# PACKING CIRCLES WITHIN CIRCULAR CONTAINERS: A NEW HEURISTIC ALGORITHM FOR THE BALANCE CONSTRAINTS CASE

Washington Alves de Oliveira[1]*, Luiz Leduino de Salles Neto[2],
Antonio Carlos Moretti[1] and Ednei Felix Reis[3]

*Corresponding author
[1]Faculdade de Ciências Aplicadas, Universidade Estadual de Campinas, 13484-350 Limeira, SP, Brazil
[2]Instituto de Ciência e Tecnologia, Universidade Federal de São Paulo, 12247-014 São José dos Campos, SP, Brazil
[3]Departamento de Matemática, Universidade Tecnológica Federal do Paraná, 84016-210 Ponta Grossa, PR, Brazil
E-mails: washington.oliveira@fca.unicamp.br [Oliveira] / luiz.leduino@unifesp.br [Salles Neto] /
moretti@ime.unicamp.br [Moretti] / edneif@utfpr.edu.br [Reis]

**ABSTRACT.** In this work we propose a heuristic algorithm for the layout optimization for disks installed in a rotating circular container. This is a unequal circle packing problem with additional balance constraints. It proved to be an NP-hard problem, which justifies heuristics methods for its resolution in larger instances. The main feature of our heuristic is based on the selection of the next circle to be placed inside the container according to the position of the system's center of mass. Our approach has been tested on a series of instances up to 55 circles and compared with the literature. Computational results show good performance in terms of solution quality and computational time for the proposed algorithm.

## 1 Introduction

We study how to install unequal disks in a rotating circular container, which is an adaptation of the model for the two-dimensional (2D) unequal circle packing problem with balance behavioral constraints. This problem arises in some engineering applications: development of satellites and rockets, multiple spindle box, rotating structure and so on. The low cost and high performance of the equipment require the best internal configuration among different geometric devices.

This problem is known as *layout optimization problem* (LOP), and consists in placing a set of circles in a circular container of minimum envelopment radius without overlap and with minimum imbalance. Each circle is characterized by its radius and mass. There, the original three-dimensional (3D) case (the equipment must rotate around its own axis) is simplified: different two-dimensional circles (see, Figure 1(c)) represent three-dimensional cylindrical objects to be placed inside the circular container.

Figure 1 illustrates the physical problem. Figure 1(a) shows a rotating cylindrical container. The symbol $\omega$ and the arrow illustrate the rotation around the axis of the equipment, $\omega$ is the angular velocity. In another viewpoint, Figure 1(b) shows the interior of the equipment where distinct circular devices need to be placed. In this example, six cylinders are placed, in which the radii, masses and heights are not necessarily equal.

Research on packing circles into a circular container has been documented and used to obtain good solutions. Heuristic, metaheuristic and hybrid methods are used in most of them. There are only a few publications discussing the disk problems with balance constraints.

LOP is a combinatorial problem and has been proved to be NP-hard (Lenstra & Rinnooy Kan 1979). This problem was first proposed by Teng et al. (1994), where a mathematical model and a series of intuitive algorithms combining the method of constructing the initial *objects topo-models* with the *model-changing iteration* method are described, and the validity of the proposed algorithms is verified by numerical examples. Tang & Teng (1999) presented a modified genetic algorithm called *decimal coded adaptive genetic algorithm* to solve the LOP. Xu et al. (2007) developed a version of genetic algorithm called *order-based positioning technique*, which finds the best ordering for placing the circles in the container, and compare it with two existing nature-inspired methods. Qian et al. (2001) extended the work (Tang & Teng 1999) by introducing a genetic algorithm based on *human-computer intervention*, in which a human expert examines the best solution obtained through the loops of many generations and designs new solutions.

Methods based on particle swarm optimization (PSO) have been applied to the LOP. Li et al. (2004) developed a PSO method *based-mutation operator*. This approach can escape from the local minima, maintaining the characteristic of fast speed of convergence. Zhou et al. (2005) proposed a hybrid approach based on *constraint handling strategy suit* for PSO, where improvement is made by using direct search to increase the local search ability of the algorithm. Xiao et al. (2007) presented two nature-inspired approaches based on *gradient search*, the first hybrid with simulated annealing (SA) method and the second hybrid with PSO method. Lei (2009) presented an adaptive PSO with a better search performance, which employs *multi-adaptive strategies* to plan large-scale space global search and refined local search to obtain global optimum.

Huang & Chen (2006) proposed an improved version of the *quasi-physical quasi-human* algorithm proposed by Wang et al. (2002) for solving the disk packing problem with equilibrium constraints. An efficient strategy of accelerating the search process is introduced in the steepest descends method to shorten the execution time. In Liu & Li (2010) the LOP is converted into an unconstrained optimization problem which is solved by the *basin filling algorithm* presented by them, together with the improved energy landscape paving method, the gradient method based on local search and the heuristic configuration update mechanism. Liu et al. (2010) presented a *simulated annealing heuristic* for solving the LOP by incorporating the neighborhood search mechanism and the adaptive gradient method. The neighborhood search mechanism avoids the disadvantage of blind search in the simulated annealing algorithm, and the adaptive gradient method is used to speed up the search for the best solution. Liu et al. (2011) developed a *tabu search algorithm* for solving the LOP. The algorithm begins with a random initial configuration and applies the gradient method with an adaptive step length to search for the minimum energy configuration. He et al. (2013) proposed a hybrid approach based on *coarse-to-fine quasi-physical* optimization method, where improvement is made by adapting the quasi-physical descent and the tabu search procedures. The algorithm approach takes into account the diversity of the search space to facilitate the global search, and it also does fine search to find the corresponding best solution in a promising local area. Liu et al. (2015) presented a heuristic based on *energy landscape paving*. The LOP is converted into the unconstrained optimization problem by using quasi-physical strategy and penalty function method. Subsequently, the heuristic approach combines a new updating mechanism of the histogram function in an improved energy landscape paving, and a local search for solving the LOP.

In this paper, we propose a new heuristic to solve the LOP. The basic idea of our approach, called *center-of-mass-based placing technique* (CMPT), is to place each circle according to the current position of the center of mass of the system.

Results for a selected set of instances are found in Huang & Chen (2006), Xiao et al. (2007), Lei (2009), Liu & Li (2010), Liu et al. (2010, 2011, 2015) and He et al. (2013). To validate our approach, we compare the results of our heuristic with these instances. Computational results show good performance in terms of solution quality and computational time.

The paper is organized as follows. Section 2 presents a formal definition of the unequal circle packing problem with balance constraints, and some definitions are established. Section 3 describes our heuristic. In Section 4, we present and analyze the experimental results, and Section 5 concludes the paper.

## 2 Problem formulation

We consider the following layout optimization for the disks installed in a rotating circular container: given a set of circles (not necessarily equal), find the minimal radius of a circular container in which all circles can be packed without overlap, and the shift of the dynamic equilibrium of the system should be minimized. The decision problem is stated as follows.

Consider a circular container of radius $r$, a set of $n$ circles $i$ of radii $r_i$ and mass $m_i$, $i \in N = \{1, \ldots, n\}$. Let $(x, y)^T$ be the coordinates of the container center, and $(x_i, y_i)^T$ the center coordinates of the circle $i$. Let $f_1(z) = r$ be the first objective function, and
$f_2(z) = \sqrt{\left(\sum_{i=1}^{n} m_i \omega^2 (x_i - x)\right)^2 + \left(\sum_{i=1}^{n} m_i \omega^2 (y_i - y)\right)^2}$ the second objective function, which measures the shift in the dynamic equilibrium of the system caused by the rotation of the container. Without loss of generality, we can consider $\omega = 1$. The problem is to determine if there exists a $(2n+3)$-dimensional vector $z = (r,x,y,x_1,y_1,x_2,y_2,\ldots,x_n,y_n)^T$ that satisfies the following mathematical formulation.

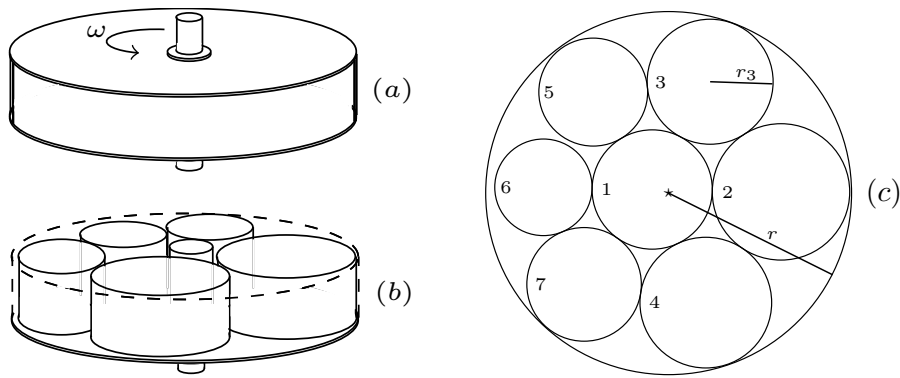

Figure 1: Circular devices inside a rotating circular container and a feasible solution

**(LOP)** Minimize $f(z) = \lambda f_1(z) + \beta f_2(z)$

subject to

$$r \geqslant \max_{1 \leqslant i \leqslant n} \left\{ r_i + \sqrt{(x_i - x)^2 + (y_i - y)^2} \right\}, \tag{1}$$

$$(x_i - x_j)^2 + (y_i - y_j)^2 \geqslant (r_i + r_j)^2, \quad i \neq j \in N, \tag{2}$$

$$z \in \mathbb{R}^{2n+3},$$

where $\lambda, \beta \in (0,1)$ are a pair of preset weights, $\lambda + \beta = 1$. Constraint (1) states that circle $i$ placed inside the container should not extend outside the container, while constraints (2) require that two any circles placed inside the container do not overlap each other.

Figure 1(c) illustrates a typical feasible solution to the LOP. The circles are numbered from 1 to 7, $r_3$ is the radius of the circle 3, there is no overlap between the circles and the seven circles are completely placed into the larger circle of radius $r$ (radius of the container).

## 2.1 Definitions

We develop a constructive heuristic guided by a simple strategy. A suboptimal solution is reached after gradually placing a circle at a time inside the container. Each circle is placed in an Euclidean coordinates system on the following evaluation criteria: select as the new position of the circle according to the current center of mass of the system without overlapping with the circles placed earlier; attempt to fill the wasted spaces after placing this circle; and in the end, select as the new coordinates of the container center that completely eliminate the dynamic imbalance of the system.

To perform the above criteria we need some notations and definitions. We denote by $X(i) = (x_i, y_i)^T$ the center coordinates of the circle $i$, by $d(i,j) = d(X(i), X(j)) = \sqrt{(x_i - x_j)^2 + (y_i - y_j)^2}$ the Euclidean distance between the center coordinates of the circles $i$ and $j$, and by $\Gamma(i,j) = \{(1-\lambda)X(i) + \lambda X(j) : 0 \leqslant \lambda \leqslant 1\}$ the set of points on the line segment whose endpoints are $X(i)$ and $X(j)$. Figure 2(a) illustrates the set $\Gamma(10,12)$.

**Definition 1.** *(Contact Pair)* If $d(i,j) = r_i + r_j$, we say that $\{i,j\}$ is a *Contact Pair* of circles.

**Definition 2.** *(Layout)* A partial *Layout*, denoted by $L$, is a partial pattern (layout) formed by a subset of the $m \geqslant 2$ of the circle centers, which have already been placed inside the container without overlap. Assume in addition that the container itself is in $L$. If $m = n$, then $L$ is a complete layout (or solution).

Figure 2(b) illustrates a partial Layout formed by 16 circles placed inside the container. Among others, $\{5,9\}$ and $\{8,15\}$ are Contact Pairs.

**Definition 3.** *(Placed Cyclic Order)* Let $C = i_1 i_2 \cdots i_{t-1} i_t$, $i_p \in N$, $p = 1, \ldots, t$, be a cyclic order of circles, which have already been placed inside the container without overlap. In addition, the intersection of any two sets $\Gamma(i_{p_1}, i_{q_1})$ and $\Gamma(i_{p_2}, i_{q_2})$ have at most one endpoint in common, $p_1, p_2, q_1, q_2 \in \{1, \ldots, t\}$. We say that $C$ is a *Placed Cyclic Order*.

**Definition 4.** *(Contact Cyclic Order)* Let $C$ be a Placed Cyclic Order. If the circles are two by two Contact Pairs in $C$, we say that $C$ is a *Contact Cyclic Order*.

Given a $C = i_1 i_2 \cdots i_{p-1} i_p i_{p+1} \cdots i_{t-1} i_t$, we say that the circles $i_1, i_2, \cdots, i_{p-1}$ are in *counterclockwise* order in relation to circle $i_p$ and the circles $i_{p+1}, \cdots, i_{t-1} i_t$ are in *clockwise* order in relation to circle $i_p$.

**Definition 5.** *(Main Area)* Let $C$ be a Placed Cyclic Order. We say that the area bounded by the union of the line segments $\Gamma(i_p, i_q)$, where $i_p, i_q$ are in $C$, is the *Main Area* of $C$, which is denoted by $A(C)$.

Figure 2(a) illustrates a Contact Cyclic Order $\bar{C}$ formed by 12 circles placed inside the container. Note that all $\{i, j\}$ in $\bar{C}$ are two by two Contact Pairs. In Figure 2(b), in addition to $\bar{C}$, it is illustrated a Contact Cyclic Order $C$ formed by 4 circles. Note that all circles in $C$ (dashed lines) are completely placed on $A(\bar{C})$. This is a feature of our approach, since several Contact Cyclic Order are obtained by circling each other. This approach is an important requirement, since it can yield a more compact layout.

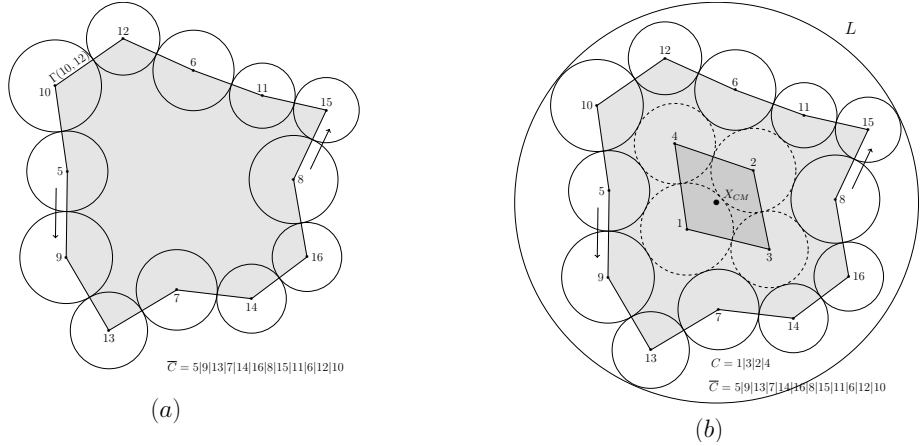

Figure 2: Two Contact Cyclic Orders and a partial Layout

Let $\bar{N} \subset N$ be a subset of circles placed inside the container and $|\bar{N}|$ be the cardinality of $\bar{N}$. We denote the *centroid* of $\bar{N}$ by the coordinates $X_C = \frac{1}{|\bar{N}|} \sum_{i \in \bar{N}} X(i)$.

**Definition 6.** *(Border)* Let $L$ be a partial Layout and $C$ be a Contact Cyclic Order. If the center of each circle in $L$ belongs to $A(C)$, we say in addition that $C$ is the *Border* of the partial Layout $L$.

Figure 2(b) illustrates the Border $\bar{C}$ (12 circles) of the partial Layout $L$ (16 circles). Note that all circle centers in $L$ belong to $A(\bar{C})$. On the other hand, $C = 1|3|2|4$ is not a Border of $L$, since $X(5) \notin A(C)$.

We consider two cases of inclusion for placing circles. In the first case, we require that the circle $k$ to be included must touch at least two previously placed circles. After this, in the second case, we require that another circle $\ell$ to be included occupies the wasted spaces after placing the circle $k$. This is a reasonable requirement, since it will generally yield a more compact layout than one defined by separate circles.

These two cases of inclusion can be explained by a partial Layout of the LOP example with seven existing circles illustrated in Figure 3. In Figure 3(a), it is shown the first case of inclusion. There are two positions to place the circle 9 (dashed lines) touching the Contact Pair $\{1,7\}$, and two positions to place the circle 6 (dashed lines) touching the Contact Pair $\{2,3\}$. Each position can be obtained by the solutions of the following particular case of the problems of Apolonio (Coxeter 1968).

$$\begin{cases} \sqrt{(x-x_{i_p})^2 + (y-y_{i_p})^2} &= r_k + r_{i_p} \\ \sqrt{(x-x_{i_q})^2 + (y-y_{i_q})^2} &= r_k + r_{i_q} \end{cases} \quad (3)$$

We denote by $St(k, i_p, i_q)$ the coordinates of the solution of the System (3) which does *not belong* to $A(C)$. Note that the System (3) has two real solutions whenever $d(i_p, i_q) \leqslant r_{i_p} + r_{i_q} + 2r_k$. In Figure 3(a), by choosing $St(9,7,1)$ as the coordinates of the circle 9, we obtain a feasible layout. However, it is not enough to choose $St(6,2,3)$ as the coordinates of the circle 6, because the circle 6 overlaps the circles 1 and 5.

In our approach, we always select the coordinates $St(k, i_p, i_q) \notin A(C)$ in order to place the new circle $k$ touching the Contact Pair $\{i_p, i_q\}$ in the Border $C$ (in Figure 3(a), we have $k = 6$ and $\{i_p, i_q\} = \{2, 3\}$), but due to the potentially large differences in the radii, it is possible to occur overlap with the circles in the Border $C$. As it is illustrated in Figure 3(a), we get around this situation by repositioning the circle $k$ to the coordinates of the solution of new System (3) for $k$, $i_{\bar{p}}$ and $i_{\bar{q}}$, where now the circle $k$ touches the circles $i_{\bar{p}}$ and $i_{\bar{q}}$ (in Figure 3(b), we have $k = 6$, $i_{\bar{p}} = 1$ and $i_{\bar{q}} = 5$). This first case of inclusion and the possible reposition define the following placement approach.

**Definition 7.** *(External Placement)* Let $L$ be a partial Layout and $C$ be the Border of $L$. An *External Placement* is the placement of a circle $k$ inside the container, so that there is no overlap, its center does not belong to $A(C)$, and it becomes Contact Pair with at least two circles in $C$. We denoted an External Placement by $p_E(k)$.

The External Placement is always selected outside $A(C)$, however if there is overlap on $C$, the repositioning of the new circle $k$ (as explained above) is done in the following routine.

**Procedure 1: External Placement routine**

**Input:** a circle $k$, a Contact Pair $\{i_p, i_q\}$, a partial Layout $L$ and the Border $C$

**Output:** an External Placement $p_E(k)$, $p$ and $q$

**Step 1.** Calculate $St(k, i_p, i_q)$ by System (3) and $p_E(k) \leftarrow St(k, i_p, i_q)$. If the circle $k$ does not overlap any circles in $C$ stop, otherwise go to Step 2.

**Step 2.** While there is overlap between the circle $k$ and the circles in $C$ repeat. If the circle $k$ overlaps the circle $i_{\bar{p}}$ furthest with respect to the counterclockwise order of the Border $C$, $p \leftarrow \bar{p}$, and if the circle $k$ overlaps the circle $i_{\bar{q}}$ furthest with respect to the clockwise order of the Border $C$, $q \leftarrow \bar{q}$, and choose $p_E(k)$ as the solution of the System (3) that is furthest from the centroid of the circles in $L$ with respect to the Euclidean distance.

---

First, if the new circle $k$ does not overlap any circles in Border $C$, the External Placement routine selects $p_E(k) = St(k, i_p, i_q)$. In our approach, this case is the most convenient way to place the next circle. However, if there is overlap, in Step 2 the routine identifies such circles ($i_{\bar{p}}$ and $i_{\bar{q}}$) in order to reposition the circle $k$ further from the centroid of the partial Layout, eventually avoiding any kind of overlap.

To obtain a more compact layout, after including the circle $k$, it is checked the possibility of including another circle to occupy the wasted spaces after placing the circle $k$. We check among the remaining circles outside the container (preferably the largest one) if there is a circle $\ell$ that can be placed into the container in a centralized position without overlap. Each centralized position is the centroid coordinates of a certain set of circles which includes the circle $k$ and the two circles touching the circle $k$. Figure 3(b) illustrates the second case of inclusion, which we can investigate the possibility of positioning a circle in the wasted space after placing the circle 9 touching the Contact Pair $\{1, 7\}$ (centroid $\hat{X}_C$ of the circles $\{1, 7, 9\}$), and in the wasted space after placing the circle 6 touching the circles 1 and 5 (centroid $\bar{X}_C$ of the circles $\{1, 2, 3, 4, 5, 6\}$).

**Definition 8.** *(Internal Placement)* Let $L$ be a partial Layout, $C$ be the Border of $L$, and $\{i_p, k\}$, $\{k, i_q\}$ be Contact Pairs in $C$, where $k$ is the previous circle included. An *Internal Placement* is the placement of a circle $\ell$ inside the container, so that its center belongs to $A(C)$, $\ell$ does not overlap with any other circle, and the center of $\ell$ is placed at the centroid coordinates of the subset $\bar{N}$, where $\{k, i_p, i_q\} \subseteq \bar{N}$. We denote an Internal Placement by $p_I(\bar{N}, \ell)$, meaning that $\ell$ is to be placed at the centroid coordinates $X_C$ of $\bar{N}$.

Let $L$ be a partial Layout and $C$ be the Border of $L$. In our algorithm, each positioning in the first case of inclusion is always done by looking at the Contact Pairs in the Border $C$. Suppose that the remaining circle $k$ is selected to be placed touching the Contact Pair $\{i_p, i_q\}$ in $C$. The placement of the circle $k$ causes the addition of one element in $L$ and one index in $C$, and perhaps the removal of some indices from $C$. This will be represented by the following operation.

$$O_-^+(k, i_p, i_{p+s})(C) = C + \{k\} - \{i_{p+1}, \ldots, i_{p+s-1}\} = i_1 i_2 \cdots i_p \, k \, i_{p+s} \cdots i_{t-1} i_t,$$

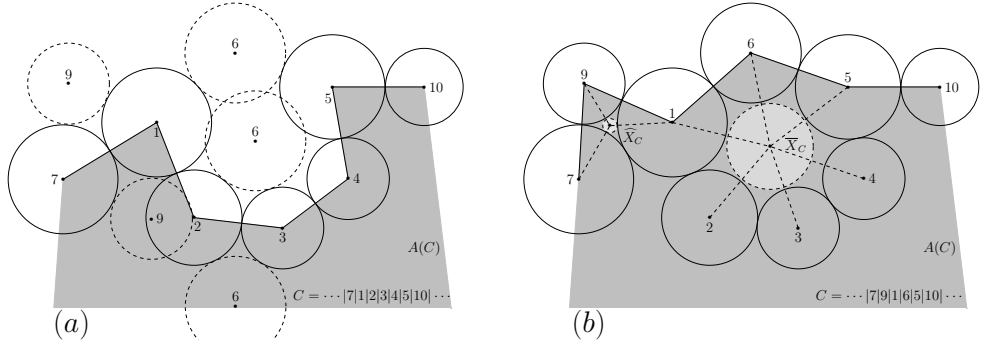

Figure 3: Two cases of inclusion: (a) External Placement and (b) Internal Placement

where $1 \leqslant s \leqslant \lfloor (t-2)/2 \rfloor$.

With this choice for $s$ there are fewer indices between $p$ and $p+s$ than $p+s$ and $p$.

The operation $O_-^+(k, i_p, i_{p+s})(C)$ applied to $C$ means that the circle $k$ was placed inside the container touching the circles $i_p$ and $i_{p+s}$ without overlap. Then the index $k$ is added to $C$, the subset of indices $\{i_{p+1}, \ldots, i_{p+s-1}\}$ between $p$ and $p+s$ is removed from $C$, and the coordinates $X(k)$ are added to the partial Layout $L$. Note that if $s=1$ there is no removal of indices from $C$, and the index $k$ is inserted in $C$ between the indices $i_p$ and $i_{p+1}$.

The possible placement of the circle $\ell$ after the placement of the circle $k$ only causes the possible addition of the coordinates $X(\ell)$ to the partial Layout $L$.

In our approach, we require the imbalance of the system be zero. It seems intuitive that this requirement may result in a good solution. We denote the center of mass of the system by $X_{CM} = (x_{CM}, y_{CM}) = (1/\sum_{i=1}^{n} m_i) \left( \sum_{i=1}^{n} m_i x_i, \sum_{i=1}^{n} m_i y_i \right)$, then one can shift the center of the rotating circular container to the center of mass of the system to have zero imbalance. This shift is made at each *outer iteration* and at the end of the algorithm, but it may increase the envelopment radius. Thus, if the Layout $L$ represents a complete solution of the LOP, we denote the radius $r$ of the container by $r = R(L) \equiv \max_{1 \leqslant k \leqslant t} \{ r_{i_k} + d(X_{CM}, X(i_k)) \}$. Moreover, the index where $R(L)$ is reached is denoted by $k_{\max} \equiv \arg(R(L))$.

## 3 Center-of-mass-based placing technique (CMPT)

We present a new placing technique which yields compact layouts and quality solutions in an efficient manner. Let $\alpha = (\alpha(1), \alpha(2), \ldots, \alpha(n))$ be a permutation of $(1, 2, \ldots, n)$. We place the circles in the partial Layout $L$ one by one according to the order defined by this permutation. Given a order of inclusion, the first circles $\alpha(1)$, $\alpha(2)$, $\alpha(3)$ and $\alpha(4)$ must be positioned as follows.

**Procedure 2: Initial layout routine**

**Input:** the circles α(1), α(2), α(3) and α(4)

**Output:** an initial Layout $L$ and the initial Border $C$

Place the circle α(1) at coordinates $X(\alpha(1)) = (0,0)$. Choose the coordinates $X(\alpha(2))$, such that α(2) touches α(1). For each circle α(3) and α(4), solve the System (3) and place them at coordinates $X(\alpha(3)) = (x_{\alpha(3)}, y_{\alpha(3)}) = St(\alpha(3), \alpha(1), \alpha(2))$ and $X(\alpha(4)) = St(\alpha(4), \alpha(1), \alpha(2))$ without overlap. $L = \{X(\alpha(1)), X(\alpha(2)), X(\alpha(3)), X(\alpha(4))\}$ and $C \leftarrow \alpha(1)|\alpha(3)|\alpha(2)|\alpha(4)$.

---

Figure 2(b) illustrates the initial $L = \{X(1), X(2), X(3), X(4)\}$ and the initial Border $C = 1|3|2|4$. Among many optional positions we can choose $X(\alpha(2))$, for example, at the *x*-axis with the coordinates $(r_{\alpha(1)} + r_{\alpha(2)}, 0)$.

Suppose we have already placed the circles $\alpha(1), \alpha(2), \ldots, \alpha(k-1)$, we describe our approach for placing the circle $\alpha(k)$, and after that, we verify the possibility of placing another circle $\alpha(\ell)$, $k < \ell \leqslant n$.

When we place the circle $\alpha(k)$ (where $k > 4$, see Procedure 2), we require that the circle touches at least two previously placed circles (see Figure 3(a) and Procedure 4). This will generally yield a more compact layout. However, we can increase the compactness of the layout if the wasted spaces after placing the circle $\alpha(k)$ can be occupied by another circle $\alpha(\ell)$ (see Figure3(b) and Procedure 4).

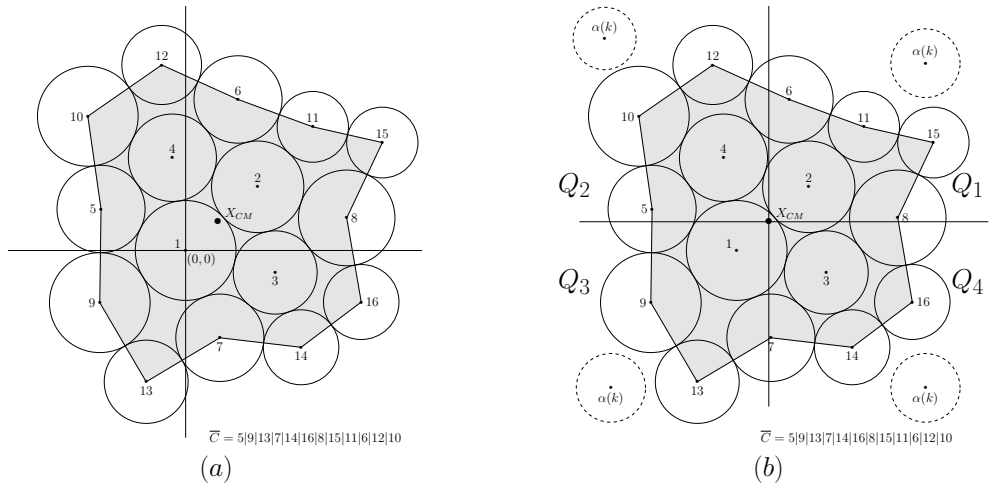

Figure 4: Example of the CMPT routine

We observe that for each additional circle, the envelopment radius of a layout is generally enlarged. In order to minimize the rate of growth of this radius during inclusions, we must properly choose a new position for circle $\alpha(k)$ which yields a smaller envelopment radius. Our strategy CMPT attempts to reduce the rate of growth of the envelopment radius by including every circle around the coordinates of the center of mass of the system, which is updated during each outer iteration.

This strategy consists of shifting the origin of the Euclidean plane to the current center of mass of the system. Then we require that the circle $\alpha(k)$ touches the circles of a Contact Pair arbitrarily chosen among the elements of the Border $C$, taking into consideration the quadrants of the Euclidean plane. This approach is performed according to the following routine.

---

**Procedure 3: CMPT routine**

**Input:** a partial Layout $L$ and the Border $C$

**Output:** the sets $Q_1$, $Q_2$, $Q_3$ and $Q_4$

**Step 1.** Calculate the coordinates of the center of mass $X_{CM}$ of the circles in $L$, and translate the origin of the Euclidean plane to $X_{CM}$.

**Step 2.** Include each Contact Pair $\{i_p, i_{p+1}\}$ of $C$ in the set $Q_h$ if the center of $i_p$ belongs to the quadrant $h$ of the Euclidean plane, for $h = 1, 2, 3, 4$.

---

Given the Border $C$, the Procedure 3 only separates the Contact Pairs in $C$ according to the quadrants of the Euclidean plane with origin shifted to the current center of mass of the system.

Figure 4 illustrates the Procedure 3. In Figure 4(a) we observe that the coordinates of the center of mass $X_{CM}$ of the system do not coincide with the coordinates of the origin $X(\alpha(1)) = (0,0)$. We wish to place the next circle $\alpha(k)$ around the coordinates $X_{CM}$ in order to mitigate the growth of the envelopment radius. By dividing the plane into quadrants, we can obtain a Border $C$ with a more circular shape. We see in Figure 4(b) that if we place each new circle $\alpha(k)$ at a different quadrant of the Euclidean plane (with the origin shifted to $X_{CM}$), then the layout is more evenly distributed.

The choice of different quadrants (a Contact Pair in $Q_h$ ($h = 1,2,3,4$)) to position the next circle $\alpha(k)$, and the operation $O_-^+$ on the Border $C$ lead to a updated Border $C$ more similar to a circular shape. This will generally yield a more compact layout, because the wasted space between the Main Area $A(C)$ and envelopment radius is minimized (see the example in Figure 5).

Next we describe the two cases of inclusion in the following routine.

---

**Procedure 4: Inclusion routine**

**Input:** a circle $\alpha(k)$, a permutation $\alpha$, the sets $Q_h$, $h = 1,2,3,4$, a Contact Pair $\{i_p, i_q\}$ in a set $Q_h$, a partial Layout $L$ and the Border $C$

**Output:** a partial Layout $L$, the Border $C$ and the sets $Q_h$, $h = 1,2,3,4$

**Step 1.** Obtain an External Placement $p_E(\alpha(k))$ and the new values for $p$ and $q$ by Procedure 1. If there are fewer indices in the Border $C$ between $q$ and $p$ than those between $p$ and $q$, then $\bar{p} \leftarrow p$, $p \leftarrow q$ and $q \leftarrow \bar{p}$.

**Step 2.** $C \leftarrow O_-^+(\alpha(k), i_p, i_q)(C)$, $X(\alpha(k)) \leftarrow p_E(\alpha(k))$, $L \leftarrow L \cup \{X(\alpha(k))\}$ and $Q_h \leftarrow Q_h \setminus \{\{i_p, i_{p+1}\}, \ldots, \{i_{q-1}, i_q\}\}$, $h = 1, 2, 3, 4$ (note that $q = p + s$, where $1 \leqslant s \leqslant \lfloor (t-2)/2 \rfloor$).

**Step 3.** If it is possible to obtain an Internal Placement $p_I(\bar{N}, \alpha(\ell))$ for the set $\bar{N} = \{\alpha(k), i_p, i_{p+1}, \ldots, i_{q-1}, i_q\}$ and a circle $\alpha(\ell)$ (preferably the largest) in the permutation $\alpha$, $k < \ell \leqslant n$, then $X(\alpha(\ell)) \leftarrow p_I(\bar{N}, \alpha(\ell))$, $L \leftarrow L \cup \{X(\alpha(\ell))\}$, and exclude $\alpha(\ell)$ from $\alpha$.

---

The Inclusion routine attempts to place the new circles in a more compact layout. First, it computes an External Placement for the next circle $\alpha(k)$ by Procedure 1 and updates the values for $p$ and $q$ in order to obtain fewer indices between $p$ and $q$ than those between $q$ and $p$. In Step 2, the Border $C$ is updated by the operation $O_-^+$, where the indices between $p$ and $q$ are removed from $C$ (by Step 1, those indices correspond to the internal circles from the Border when compared to the indices between $q$ and $p$), and $\alpha(k)$ is added to $C$. The circle $\alpha(k)$ is placed inside the container and all Contact Pairs between $p$ and $q$ (including $\{i_p, i_{p+1}\}$ and $\{i_{q-1}, i_q\}$) are removed from the sets $Q_h$, $h = 1, 2, 3, 4$. This choice guarantees that the operation $O_-^+$ will exclude the internal circles from the Border $C$. Finally, a search to place another circle $\alpha(\ell)$ (Internal Placement) is performed.

*Post-optimization*

A post-optimization is performed after the algorithm builds a complete solution (represented by Layout $L$), which contemplates improvements via circle repositioning at the Border $C$ of $L$. This post-optimization process causes changes in $C$, where an index is removed and then it is repositioned in $C$ by operation $O_-^+$. The removal of the index from $C$ will be represented by the following operation.

$$D(i_p)(C) = i_1 i_2 \cdots i_{p-1}\, i_{p+1} \cdots i_{t-1} i_t.$$

The operation $D(i_p)(C)$ applied to $C$ means that the circle $i_p$ is deleted from its position. We delete the current $X(i_p)$ from the Layout $L$, and we test if a new position $p_E(i_p)$ for $i_p$ improves the radius $R(L)$ of the container.

*Main routine*

We choose to position each circle inside the container according to the following main procedure.

---

**Main routine**

**Input:** a permutation $\alpha = (\alpha(1), \alpha(2), \ldots, \alpha(n))$

**Output:** a Layout $L$ (complete solution)

**Step 1.** (*Initialization*) Obtain the initial Layout $L$ and the initial Border $C$ by Procedure 2, $k \leftarrow 5$.

**Step 2.** (*CMPT*) Obtain the sets $Q_h$, $h = 1, 2, 3, 4$ by Procedure 3.

**Step 3.** (*Layout construction*) While there are Contact Pairs in any $Q_h$ and circles outside the container, repeat for each $h = 1, 2, 3, 4$.

> If $Q_h \neq \emptyset$, choose an arbitrary $\{i_p, i_q\} \in Q_h$ and include the circle $\alpha(k)$ and the possible circle $\alpha(\ell)$, $k < \ell \leqslant n$, by Procedure 4 and $k \leftarrow k+1$.

**Step 4.** If there are circles outside the container, return to Step 2. Otherwise, go to Step 5.

**Step 5.** (*Post-optimization*) $\bar{L} \leftarrow L$, $\bar{C} \leftarrow C$, $\bar{r} \leftarrow R(\bar{L})$, compute $k_{\max}$ in $\bar{C}$ and $k \leftarrow k_{\max}$.

**Step 5.1.** $(\bar{x}, \bar{y}) \leftarrow X(i_k)$, delete the current $X(i_k)$ and repeat Step 5.2. for each Contact Pair $\{i_p, i_q\}$ of $\bar{C}$, excluding $\{i_{k-1}, i_k\}$ and $\{i_k, i_{k+1}\}$.

**Step 5.2.** Obtain an External Placement $p_E(i_k)$ and the new values for $p$ and $q$ by Procedure 1, and $X(i_k) \leftarrow p_E(i_k)$. If the radius of the container is improved, then $\bar{C} \leftarrow D(i_k)(\bar{C})$, $\bar{C} \leftarrow O_-^+(i_k, i_p, i_q)(\bar{C})$, $L \leftarrow \bar{L}$, $C \leftarrow \bar{C}$ and return to Step 5.

**Step 5.3.** $X(i_k) \leftarrow (\bar{x}, \bar{y})$ and finish the routine with the complete solution $L \leftarrow \bar{L}$, whose container center is the center of mass of the system.

---

Given a permutation $\alpha$, the Main routine builds an initial Layout $L$ in Step 1 by placing the first four circles as in Procedure 2. Next, in Step 2 the main aspect of our approach is performed by Procedure 3 (CMPT routine), where the Euclidean plane is divided into four parts and the subsets $Q_h$ ($h = 1, 2, 3, 4$) of Contact Pairs are obtained. Next, Step 3 is repeated by looking at each subset $Q_h$ and while there are circles remaining to be placed. In this step, an arbitrary Contact Pair in $Q_h$ is chosen and the two cases of inclusions are performed by Procedure 4. After we finish placing all circles inside the container, we obtain a complete solution $L$ and its Border $C$. Then, in Step 5, a post-optimization is performed via circle repositioning at the Border $C$, which attempts improvements in the envelopment radius. In the end, the center of the container is shifted to center of mass of the system, which achieves zero imbalance.

*Order of placement of the circles*

As previously described, a permutation $\alpha = (\alpha(1), \alpha(2), \ldots, \alpha(n))$ of $(1, 2, \ldots, n)$ is used as an input in our algorithm to generate a layout by specifying the order in which the circles are placed. Since there exist $n!$ possible permutations for $n$ circles, we need an appropriate technique in order to search in such a large space. Preliminary tests show that the wasted spaces after placing circles are minimized with greater efficiency when the order of addition of the circles favors those of larger radii.

Let $\alpha = (\alpha(1), \alpha(2), \ldots, \alpha(n))$ be a sequence obtained by considering their radii in descending order of the circles, i.e., $r_{\alpha(k)} \geqslant r_{\alpha(j)}$, $1 \leqslant k < j \leqslant n$. Choose an integer $b$, $1 \leqslant b \leqslant n$ and subdivide the terms of the sequence $\alpha$ in $\ell = \lfloor n/b \rfloor$ blocks. Thus, it is possible to obtain a subsequence $\bar{\alpha}$ of $\alpha$ to be used as an input to the algorithm, by permuting the positions of the first $\alpha(1), \ldots, \alpha(\ell)$ elements of $\alpha$, the $\alpha(\ell+1), \ldots, \alpha(2\ell)$ elements of $\alpha$, and so on, until we permute the positions of $\alpha(b\ell), \ldots, \alpha(n-1), \alpha(n)$ last elements of $\alpha$. With this procedure, several subsequences to place the different circles may be generated. Actually, there are $((\ell!)^b)(n-(b\ell))!$ possibilities, so that $1 \leqslant ((\ell!)^b)(n-(b\ell))! \leqslant n!$. Thus, when $b = n$, we only obtain the sequence $\bar{\alpha} = \alpha$, and when $b = 1$, we can generate at most $n!$ distinct subsequences. In our numerical experiments, for each instance of dimension greater than or equal to 10 we chose $b = 5$ to generate such sequences.

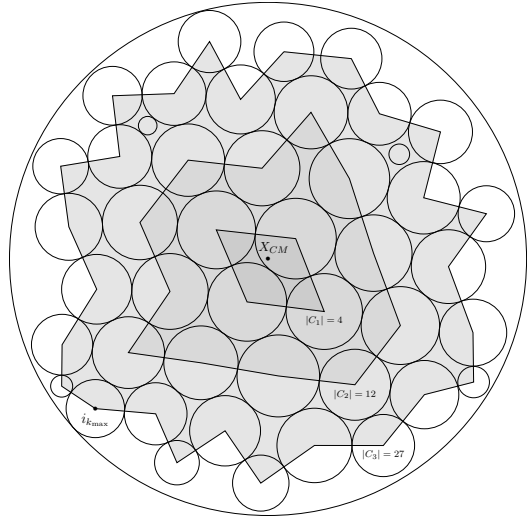

Figure 5: A suboptimal solution: 45 circles inside the container

*Complexity*

The analysis of the real computational time of the Main routine is difficult, because it does not depend only on the number of circles, but also on the diversity of the circle radii and the number of circles in a current Border $C$, as well as the implementation. Here, we analyze the upper bound of the complexity of the Main routine, when it finds a complete solution $L$ with Border $C$, such that $|C| = \lambda n$, where $0 < \lambda \leqslant 1$, including the post-optimization process. Recall that, before post-optimization, the circles in Border $C$ are two by two Contact Pair.

Given a partial Layout $L$ with $m$ circles already placed inside the container and $n - m$ circles outside. Let $|C|$ be the number of circles in the Border $C$ of the partial Layout $L$.

The strategy CMPT in Procedure 3 checks the position of $|C|$ circles in the Euclidean plane, which is done in $O(|C|)$.

When we position the circle $i_k$ (where $k > 2$, see Procedure 2) touching two circles in Border $C$, $|C|$ existing circles can define $2 \times |C|$ positions, since two existing circles define two possible positions for the third $i_k$. To determine an External Placement for $i_k$, we must check the overlap with $A(C)$ or with any circles in $C$. This is the same that we check the overlap with each circle in $L$, that is, $m$ circles (a good implementation can reduce the number of checks). Because we assess $2 \times |C|$ positions when we place the circle $i_k$, each time checking for overlaps $m$ times, then the complexity to obtain an External Placement is about $2 \times m \times |C|$.

After placing the circle $i_k$, we must check if there is a circle $i_\ell$ outside the container to be placed in an Internal Placement, then we must check the overlaps among $n-m$ circles and a subset $\bar{N}$ in $C$, which is done in about $|\bar{N}| \times (n-m) \leqslant |C| \times (n-m)$.

In the post-optimization process, we select one circle in $C$ and assess at most $(|C|-2)$ Contact Pairs in $C$ to try to improve of the envelopment radius by checking at most $2 \times (|C|-2)$ External Placements, thus the complexity of the post-optimization is bounded by $2 \times (n-1) \times (|C|-2)$.

Therefore the complexity of placing $n$ circles during the Main routine is bounded by $O(n^2|C|)$. After placing the new circle $i_k$, the operation $O_-^+$ modifies the Border $C$. This operation controls the size of $C$ during the iterations. Since $|C| \leqslant n$, the theoretical upper bound is $O(n^3)$.

# 4  Experimental results

In this section, we measure the quality and performance of our algorithm on a series of instances up to 55 circles from literature. We tested three sets of instances from the literature. We compare our approach with a series of hybrid nature-inspired approaches based on simulated annealing and particle swarm optimization (Xiao et al. 2007, Lei 2009), a hybrid approach based on simulated annealing, neighborhood search mechanism and the adaptive gradient method Liu et al. (2010), a hybrid tabu search algorithm and gradient method Liu et al. (2011), a series of heuristics based on energy landscape paving, gradient method and local search Liu & Li (2010), Liu et al. (2010, 2015), and a series of algorithms based on quasi-physical approaches, gradient method and local search Huang & Chen (2006), He et al. (2013), Liu et al. (2015).

Table 1: Data of each instance

First set of instances (Liu & Li 2010)

| Size | Radii | Mass | Size | Radii | Mass |
|---|---|---|---|---|---|
| 5 | [20.71, 50] | [20.71, 50] | 7 | [8.5, 12] | [72.25, 144] |
| 9 | [0.17, 0.41] | [0.17, 0.41] | 40 | [81, 120] | [6, 14] |
| 5 | [6, 11] | [36, 121] | | | |

Second set of instances (Huang & Chen 2006)

| Size | Radii | Mass | Size | Radii | Mass |
|---|---|---|---|---|---|
| 7 | [20, 20] | [400, 400] | 17 | [5, 25] | [25, 625] |
| 12 | [23.72, 100] | [562.64, $10^4$] | 37 | [20, 20] | [400, 400] |
| 15 | [1, 15] | [1, 225] | 50 | [20, 20] | [400, 400] |

Third of instances (Xiao et al. 2007)

| Size | Radii | Mass | Size | Radii | Mass |
|---|---|---|---|---|---|
| 10 | [5, 23] | [20, 93] | 35 | [7, 24] | [10, 99] |
| 15 | [6, 24] | [12, 98] | 40 | [6, 23] | [12, 99] |
| 20 | [5, 24] | [11, 94] | 45 | [6, 24] | [11, 99] |
| 25 | [6, 24] | [11, 96] | 50 | [5, 24] | [10, 99] |
| 30 | [6, 24] | [12, 97] | 55 | [6, 24] | [13, 99] |

These methods search for the optimal layout by directly evolving the positions of every circle, as well as considering imbalance. We use the benchmark suite of 21 instances of the problem described in Table 1 to test our algorithm. For each instance we present the range for $r_i$ and $m_i$. A more detailed description of the instances can be found in Huang & Chen (2006), Liu & Li (2010) and Xiao et al. (2007).

Table 2: Numerical results for the first set of instances

| Instance | $n$ | Huang & Chen (2006) | | | Lei (2009) | | | Liu & Li (2010) | | |
|---|---|---|---|---|---|---|---|---|---|---|
| | | $f_1$ | $f_2$ | time(s) | $f_1$ | $f_2$ | time(s) | $f_1$ | $f_2$ | time(s) |
| 1.1 | 5 | 120.710 | $7.63\times10^{-4}$ | 3.82 | – | – | – | 120.710678 | $3.89\times10^{-9}$ | 0.594 |
| 1.2 | 9 | – | – | – | – | – | – | **1.000000** | $9.7\times10^{-9}$ | 2.891 |
| 1.3 | 5 | 22.327 | $3.90\times10^{-4}$ | 2.91 | – | – | – | 22.249000 | $3.12\times10^{-9}$ | 1.313 |
| 1.4 | 7 | 31.981 | $4.11\times10^{-3}$ | 76.33 | 31.924 | $1.40\times10^{-5}$ | 427 | 31.854600 | $1.5\times10^{-9}$ | 54.828 |
| 1.5 | 40 | 742.750 | $5.40\times10^{-4}$ | 12.27 | 769.819 | $3.25\times10^{-4}$ | 1724 | 725.043500 | $2.47\times10^{-9}$ | 7.187 |
| Instance | $n$ | | Liu et al. (2010) | | | | Liu et al. (2011) | | | |
| | | | $f_1$ | $f_2$ | time(s) | | $f_1$ | $f_2$ | times(s) | |
| 1.1 | 5 | | 120.7106782 | $4.30\times10^{-9}$ | 0.969 | | 120.710678 | $3.25\times10^{-9}$ | 0.350 | |
| 1.2 | 9 | | – | – | – | | **1.000000** | $2.38\times10^{-9}$ | 0.225 | |
| 1.3 | 5 | | 22.2463 | $3.52\times10^{-9}$ | 0.469 | | **22.246203** | $2.63\times10^{-9}$ | 0.506 | |
| 1.4 | 7 | | 31.8412 | $2.03\times10^{-10}$ | 58.797 | | 31.841134 | $5.89\times10^{-9}$ | 27.766 | |
| 1.5 | 40 | | 716.6782 | $2.86\times10^{-9}$ | 154.594 | | 716.610932 | $1.77\times10^{-9}$ | 63.263 | |
| Instance | $n$ | He et al. (2013) | | | Liu et al. (2015) | | | Our algorithm | | |
| | | $f_1$ | $f_2$ | time(s) | $f_1$ | $f_2$ | time(s) | $f_1$ | $f_2$ | time(s) |
| 1.1 | 5 | 120.710678 | $2.11\times10^{-10}$ | 0.046 | 120.710678 | $3.49\times10^{-10}$ | – | **120.710618** | 0 | 1.925 |
| 1.2 | 9 | **1.000000** | $2.00\times10^{-12}$ | 0.172 | **1.000000** | $9.92\times10^{-10}$ | – | **1.000000** | 0 | 2.365 |
| 1.3 | 5 | 22.246203 | $3.44\times10^{-10}$ | 0.015 | 22.246203 | $9.22\times10^{-10}$ | – | 23.624138 | 0 | 1.812 |
| 1.4 | 7 | 31.841134 | $2.90\times10^{-10}$ | 98.031 | **31.841133** | $1.13\times10^{-9}$ | – | 31.905109 | 0 | 110.912 |
| 1.5 | 40 | **709.812500** | $1.85\times10^{-10}$ | 434.734 | 715.978266 | $3.15\times10^{-7}$ | – | 745.118349 | 0 | 829.156 |

The routines were implemented in MATLAB language, and executed on a PC with an Intel Core i7, 7.7 GB of RAM and Linux operating system.

Except for the instances 1.1 and 1.3, both of them with 5 circles, we decide to generate 7! = 5040 distinct permutations α as input for the algorithm in each instance, i.e., we fixed in 5040 the number of executions of the Main routine for each instance and the best solution found was selected. This amount of tests proved adequate for our comparisons.

The results from the first, second and third sets of instances are presented in Tables 2, 3 and 4 respectively, where we compare our approach with those described in each indicated reference. The results are shown for the size of the instances, the best radius of the container obtained (first objective function $f_1$), the imbalance obtained (second objective function $f_2$), and the running time $t$ (in seconds).

Table 2 shows that our approach proved to be competitive. We obtained the best value for $f_1$ on instance 1.1, tied in instance 1.2, while obtained results 6,1%, 0,2% and 4,0% worse than the best result on instances 1.3, 1.4 and 1.5 respectively.

Similar results were obtained in the second set test, as can be seen in Table 3. Our algorithm tied in instance 2.1 and 2.2, while obtained results at most 7,7% worse than the best result on instances 2.3, 2.4, 2.5 and 2.6.

In Table 4 we compare our approach with three other algorithms. The first set of data are from a version of simulated annealing (SA), the second set of data are from the same reference, but one of them is a version of particle swarm optimization (PSO), while the third set of data is a heuristic based on energy landscape paving. Again, our approach proved to be competitive. In relation to the envelopment radius we obtained better results in 5 out of the 10 instances. We only obtained worse results in five cases, but they were on average approximately 0.63% worse than the results from literature for such instances.

Table 3: Numerical results for the second set of instances

| Instance | n | Huang & Chen (2006) | | | Liu et al. (2010) | | | Liu et al. (2011) | | |
|---|---|---|---|---|---|---|---|---|---|---|
| | | $f_1$ | $f_2$ | time(s) | $f_1$ | $f_2$ | time(s) | $f_1$ | $f_2$ | time(s) |
| 2.1 | 7 | 60.00 | $3.6 \times 10^{-3}$ | 7.78 | 60.0000 | $1.59 \times 10^{-9}$ | 1.031 | **60.000000** | $7.38 \times 10^{-9}$ | 0.206 |
| 2.2 | 12 | 215.470 | $9.5 \times 10^{-3}$ | 444.78 | **215.470054** | $2.74 \times 10^{-7}$ | 11.562 | **215.470054** | $3.77 \times 10^{-7}$ | 3.042 |
| 2.3 | 15 | 39.780 | $7.6 \times 10^{-3}$ | 91.92 | 39.123400 | $2.97 \times 10^{-8}$ | 130.362 | 39.065105 | $4.13 \times 10^{-8}$ | 365.797 |
| 2.4 | 17 | 49.720 | $5.1 \times 10^{-3}$ | 157.92 | 49.507800 | $6.83 \times 10^{-8}$ | 123.407 | 49.368194 | $2.50 \times 10^{-8}$ | 597.719 |
| 2.5 | 37 | 135.176 | $6.7 \times 10^{-3}$ | 18.29 | 135.175410 | $1.23 \times 10^{-8}$ | 0.203 | **135.175410** | $4.91 \times 10^{-9}$ | 0.425 |
| 2.6 | 50 | 159.570 | $8.0 \times 10^{-3}$ | 348.97 | 158.967800 | $4.81 \times 10^{-9}$ | 453.625 | 158.967698 | $3.49 \times 10^{-9}$ | 1078.386 |
| Instance | n | He et al. (2013) | | | Liu et al. (2015) | | | Our algorithm | | |
| | | $f_1$ | $f_2$ | time(s) | $f_1$ | $f_2$ | time(s) | $f_1$ | $f_2$ | time(s) |
| 2.1 | 7 | **60.000000** | $2.66 \times 10^{-9}$ | 0.015 | **60.000000** | $5.13 \times 10^{-9}$ | – | **60.000000** | 0 | 0.118 |
| 2.2 | 12 | **215.470054** | $2.92 \times 10^{-8}$ | 0.391 | **215.470054** | $1.69 \times 10^{-7}$ | – | **215.470054** | 0 | 52.125 |
| 2.3 | 15 | **39.037635** | $7.78 \times 10^{-10}$ | 36.343 | 39.065105 | $1.12 \times 10^{-8}$ | – | 40.090538 | 0 | 167.248 |
| 2.4 | 17 | **49.338750** | $1.51 \times 10^{-9}$ | 10.562 | 49.368194 | $1.42 \times 10^{-8}$ | – | 50.246873 | 0 | 325.080 |
| 2.5 | 37 | 135.175410 | $6.95 \times 10^{-9}$ | 0.234 | **135.175410** | $3.35 \times 10^{-9}$ | – | 145.624231 | 0 | 0.402 |
| 2.6 | 50 | **158.963672** | $5.14 \times 10^{-9}$ | 268.078 | 158.967698 | $3.65 \times 10^{-9}$ | – | 163.206698 | 0 | 0.390 |

Overall, running time obtained by our algorithm can be considered good. Since the center of the rotating circular container is shifted to the center of mass of the system we always have $f_2 = 0$, making our solutions more interesting than the others for this first set of instances.

Figure 5 illustrates a typical solution obtained by our algorithm for an instance of 45 circles. Note that the large Border $C_3$ have 27 circles, i.e., 60% of the size, and when we carefully read the CMPT routine, we can see that the initial Border $C_1$ ($|C_1| = 4$) is iteratively transformed in the Border $C_2$ ($|C_2| = 12$), and finally the latter is iteratively transformed in the Border $C_3$. In this example there were only two inclusions by Internal Placement.

The computational results show that the proposed algorithm is an effective method for solving the circular packing problem with additional balance constraints.

# 5 Conclusions

We have presented a new heuristic called center-of-mass-based placing technique for packing unequal circles into a 2D circular container with additional balance constraints. The main feature of our algorithm is the use of the Euclidean plane with origin in the center of mass of the system to select a new circle to be placed inside the container. We evaluate our approach on a series of instances from the literature and compare with existing algorithms. The computational results show that our approach is competitive and outperforms some published methods for solving this problem. We conclude that our approach is simple, but with high performance. Future work will focus on the problem of packing spheres.

Table 4: Numerical results for the third set of instances

| Instance | n | Xiao et al. (2007) (SA) | | | Xiao et al. (2007) (PSO) | | |
|---|---|---|---|---|---|---|---|
| | | $f_1$ | $f_2$ | time(s) | $f_1$ | $f_2$ | time(s) |
| 3.1 | 10 | 60.96 | 0 | 3237 | 59.93 | 0 | 2898 |
| 3.2 | 15 | 68.77 | 0 | 8320 | 67.65 | 0 | 8659 |
| 3.3 | 20 | 83.09 | 0 | 18431 | 83.06 | 0 | 20035 |
| 3.4 | 25 | 83.97 | 0 | 34032 | 84.24 | 0 | 36815 |
| 3.5 | 30 | 99.58 | 0 | 54565 | 99.89 | 0 | 62360 |
| 3.6 | 35 | 102.86 | 0 | 76760 | 102.71 | 0 | 86537 |
| 3.7 | 40 | 115.15 | 0 | 128112 | 115.58 | 0 | 122390 |
| 3.8 | 45 | 120.63 | 0 | 167484 | 119.67 | 0 | 153006 |
| 3.9 | 50 | 125.82 | 0 | 198071 | 126.19 | 0 | 199050 |
| 3.10 | 55 | 138.22 | 0 | 198071 | 138.89 | 0 | 244171 |
| Instance | n | Liu et al. (2015) | | | Our algorithm | | |
| | | $f_1$ | $f_2$ | time(s) | $f_1$ | $f_2$ | time(s) |
| 3.1 | 10 | **59.92** | $5.40\times10^{-9}$ | – | 59.96 | 0 | 28 |
| 3.2 | 15 | **67.39** | $6.44\times10^{-9}$ | – | 67.59 | 0 | 39 |
| 3.3 | 20 | **82.99** | $2.48\times10^{-9}$ | – | 83.85 | 0 | 82 |
| 3.4 | 25 | **82.98** | $5.10\times10^{-9}$ | – | 83.68 | 0 | 107 |
| 3.5 | 30 | 98.97 | $4.14\times10^{-9}$ | – | **98.95** | 0 | 134 |
| 3.6 | 35 | 102.32 | $4.50\times10^{-9}$ | – | **102.15** | 0 | 155 |
| 3.7 | 40 | 115.00 | $5.82\times10^{-9}$ | – | **114.15** | 0 | 96 |
| 3.8 | 45 | 119.07 | $7.64\times10^{-9}$ | – | **117.69** | 0 | 118 |
| 3.9 | 50 | 124.98 | $4.45\times10^{-9}$ | – | **124.32** | 0 | 136 |
| 3.10 | 55 | **136.13** | $2.19\times10^{-9}$ | – | 137.37 | 0 | 154 |

# Acknowledgements

The authors are indebted to the anonymous reviewers for their helpful comments. The first author wishes to thank CAPES and FAEPEX-UNICAMP (grant 285/15), the second author is grateful to FAPESP, the third author thank CNPq.